# Recent Applications of Optical Parametric Amplifiers in Hybrid WDM/TDM Local Area Optical Networks

**Abd El–Naser A. Mohamed[1], Mohamed M. E. El-Halawany[2]**
**Ahmed Nabih Zaki Rashed[3*] and Mahmoud M. A. Eid[4]**
**[1,2,3,4]Electronics and Electrical Communication Engineering Department**
**Faculty of Electronic Engineering, Menouf 32951, Menoufia University, EGYPT**
**[1]E-mail: Abd_elnaser6@yahoo.com, [3*]E-mail: ahmed_733@yahoo.com**
**Tel.: +2 048-3660-617, Fax: +2 048-3660-617**

***Abstract*—In the present paper, the recent applications of optical parametric amplifiers (OPAs) in hybrid wavelength division multiplexing (WDM)/time division multiplexing (TDM) local area passive optical networks have been modeled and parametrically investigated over wide range of the affecting parameters. Moreover, we have analyzed the ability of the hybrid WDM/TDM Passive optical networks to handle a triple play solution, offering voice, video, and data services to the multiple users. Finally, we have investigated the maximum time division multiplexing (MTDM) bit rates for optical network units (ONUs) for maximum number of supported users with optical parametric amplifier technique across the single mode fiber (SMF) or highly nonlinear fiber (HNLF) cables to achieve both maximum network reach and quality of service (QOS).**

## I. INTRODUCTION

Optical access networks present the future-proof alternative to the currently deployed copper access infrastructure [1]. With the standardization of time-division-multiplexing passive optical networks (TDM-PONs), a cost-effective access technology based on optics has been developed [2]. However, further development needs to be carried out in order to fully exploit the benefits of optical fiber technology. WDM-PONs are an option, where capacity per user can be very high, but their cost does not make them attractive for practical implementation nowadays. Several recent proposals have demonstrated the feasibility of combining WDM and TDM to optimize network performance and resource utilization. PONs WDM technology (such as WDM PON and hybrid WDM/TDM-PON can provide a promising solution for broadband access [3]. High utilization of wavelengths is desirable to support more subscribers in access networks since the total number of wavelengths (offered by commercially available light sources) is limited. Furthermore, access networks are very cost-sensitive. Because network operators need to guarantee the level of connection availability specified in the service level agreement it is important in PON deployment to minimize the cost for protection while maintaining the connection availability at the acceptable level [4]. Typically, PON consists of optical line terminal (OLT), remote node (RN), several ONUs, and fiber links including feeder fiber between OLT and RN which are shared by all the ONUs and distribution fibers (DFs) between RN and each ONU [5]. Obviously, multiplying these network resources (and investment cost) to provide protection is not acceptable in access networks. Therefore, much effort has been made to develop cost-effective protection schemes for PONs [6]. Fiber interconnection between two neighboring ONUs is used to provide the protection for DFs which allows us to save a lot of investment cost. The protection architecture presented in [7] requires only half of the wavelengths as compared with , but needs much more interconnecting fiber between the ONUs. With the explosive growth of end user demand for higher bandwidth, various types of PONs have been proposed. PON can be roughly divided into two categories such as TDM and WDM methods. Compared with TDM-PONs, WDM-PON systems allocate a separate wavelength to each subscriber, enabling the delivery of dedicated bandwidth per ONU. Moreover, this virtual point-to-point connection enables a large guaranteed bandwidth, protocol transparency, high quality of service, excellent security, bit-rate independence, and easy upgradeability. Especially, recent good progress on a thermal arrayed waveguide grating (AWG) and cost-effective colorless ONUs has empowered WDM-PON as an optimum solution for the access network. However, fiber link failure from the OLT to the ONU leads to the enormous loss of data [8].

Parametric amplification is a well-known phenomenon in materials providing $\chi^{(2)}$ nonlinearity. However, parametric amplification can also be obtained in optical fibers exploiting the $\chi^{(3)}$ nonlinearity.

New high-power light sources and optical fibers with a nonlinear parameter 5–10 times higher than for conventional fibers , as well as the need of amplification outside the conventional Erbium band has increased the interest in such OPAs. The fiber-based OPA is a well-known technique offering discrete or lumped gain using only a few hundred meters of fiber [9]. It offers a wide gain bandwidth and may in similarity with the Raman amplifier be tailored to operate at any wavelength. An OPA is pumped with one or several intense pump waves providing gain over two wavelength bands surrounding the single pump wave, or in the latter case, the wavelength bands surrounding each of the pumps [10]. As the parametric gain process do not rely on energy transitions between energy states it enable a wideband and flat gain profile contrary to the Raman and the Erbium-doped fiber amplifier (EDFA). OPAs have been intensively studied in recent years due to their potential use for amplification and wavelength conversion in multi-terabit/sec. WDM transmission systems [11]. OPAs have the advantage of being able to operate in any of the telecom bands (S–C–L) depending upon pump wavelength and the fiber zero dispersion wavelength, which, in principle, can be appropriately tailored. In order to be a practical amplifier in a WDM system, the OPA should exhibit high gain, large bandwidth and should be spectrally flat, among other requirements [12]. OPAs have been used in several applications, e.g., as wavelength converters, amplifiers, and pulse sources. Based on four wave mixing (FWM) with exponential gain [13], the OPA can (when used as a pulse source) generate short return-to-zero (RZ) pulses at the input signal wavelength and the converted idler wavelength [14]. This is due to the fact that the peak of the pump pulse, which has higher gain than its wings, results in a signal pump pulse compressed with respect to the total pump pulse [15].

In the present study, the OPAs are deeply studied and parametrically investigated over wide range of the affecting parameters in hybrid WDM/TDM local area Passive optical networks across SMF, or HNLF cables to achieve the best QOS to handle a triple play solution (video, voice, and data) to the supported users.

## II. SIMPLIFIED HYBRID WDM/TDM LOCAL AREA PASSIVE OPTICAL NETWORK ARCHITECTURE MODEL

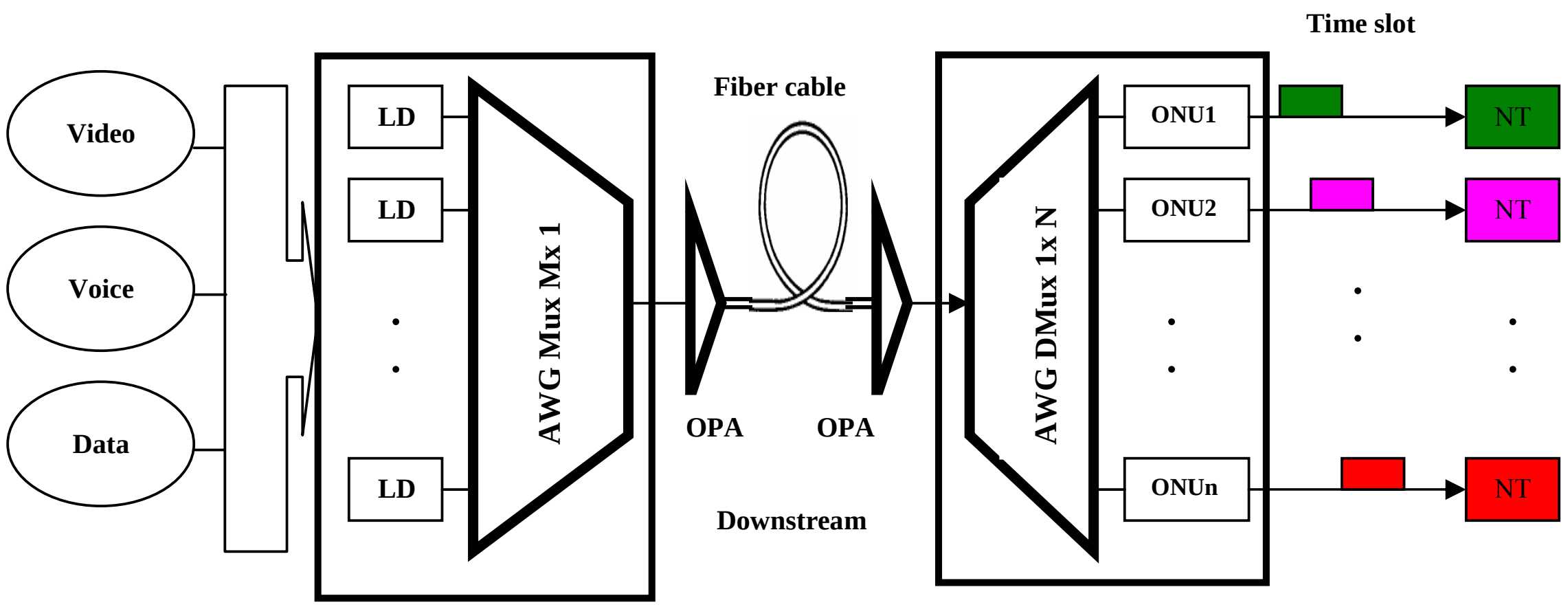

Figure1. Hybrid WDM/TDM Local Area PON Architecture Model

WDM/TDM local area PON is considered as a compromise between WDM-PON and TDM-PON which combines the advantages of both technologies [6]. The architecture model of WDM/TDM PON is shown in Fig. 1. WDM/TDM PON consists of many laser diodes as a source of optical signals, arrayed waveguide grating multiplexer (AWG Mux) in the OLT, optical fiber cable, two optical parametric amplifiers to strength the optical signal wavelength, arrayed waveguide grating demultiplexer (AWG Demux), ONU in the RN, and optical time division multiplexer (OTDM) which lies in the ONU and network terminal (NT) which connects to the user. In the downstream direction, traffic including video, voice, and data is transmitted from the backbone network to the OLT and according to different users and location, data is transmitted into corresponding wavelength and multiplexed by AWG Mux. When traffic arrives at RN, wavelengths are demultiplexed by AWG Demux and sent to different fibers. Each fiber (wavelength) serves several NTs. Signal in each wavelength is demultiplexed by OTDM and different time slots are sent to corresponding users [7]. TDM PON has emerged as a promising technology to replace conventional access network. To put it simple, PON consists of OLT which lies in the central office, passive optical splitter and ONU which lies at the users. Unlike digital subscriber line (DSL), this is a point to multipoint topology without any active components from the central office to the users. The TDM scheme reduces cost and provide a very efficient method since several users will share the same wavelength. But it also brings some problems such as security issues: because of it broadcast nature and the truth that many users share the bandwidth, transmission distance between user and OLT is limited due to the fact that many users share an optical splitter, and the protocol needed to implement TDM and dynamically allocate bandwidth is very complicated and not easy to realize. Because of these disadvantages, WDM-PON is proposed currently. In the WDM-PON, for each user there is a dedicated wavelength from OLT to ONU. Obviously, this is a point to point topology which differs from point to multipoint. The optical splitter is replaced by RN which actually is a multiplexer/demultiplexer and each user is equipped with a transmitter and receiver. In the downstream direction, optical signals designated for different users are transmitted in their own dedicated wavelength from the OLT and multiplexed in a single fiber cable. At the receiver side which is the RN, the demultiplexer demultiplexes the wavelengths and the signals will be received respectively by each user. But the only problem which is a very big issue of WDM-PON is the low efficiency and high cost [8]. Gbit/sec rates of WDM-PON is too large for a single user so most of the time, big portion of the bandwidth of one wavelength is wasted. Also, due to large number of wavelengths needed in WDM-PON, more fibers will be employed and more transceivers, the cost needed to build such architecture as well as the maintenance fee, all of these will add the cost for each user to afford. Thus, unless the big drop of components and component installation fee and also large increase of bandwidth demand, it is actually impractical to implement WDM-PON in a least a few years. So the directions to hybrid WDM/TDM PON is urgent to achieve the advantages of both technologies.

## III. BASIC MODEL AND ANALYSIS

Considering the minimum bandwidth per user that would be offered in a saturated case, where optical network units (ONUs) transmit at their maximum capacity [12]:

$$BW_{user} = \frac{K\,d\,T}{N\,M\left(d\,T + T_{Laser}\right)} = \frac{d\,T}{T_{window}} \quad , \qquad (1)$$

$$T_{window} = \frac{N\,M}{K}\left(d\,T + T_{Laser}\right) \quad , \qquad (2)$$

where K is the number of lasers at the OLT (typically K=M), d is the data rate, N is the number of of input ports of distribution AWG Mux, M is the number of output ports of distribution AWG Demux, T is the time slot assigned to each ONU in time units, and $T_{window}$ is maximm delay. In order to reduce the effect of $T_{Laser}$, it is very clear that the solution is to increase T. However, in that case the interval of service to serve the same ONU may be too wide for a certain applications. Therefore, a compromise must be met combining $BW_{user}$ and delay parameters. Equation (2) has been developed supposing a deterministic situation where the users are transmitting at full rate under TDM conditions. As we need to switch to all active ONUs on the network segment, there is a delay between the packet generation and the moment when the system is prepared to process it, it defined as the following expression [12]:

$$T_{window} = \rho \frac{W}{2}\left(T_{tx} + T_{Laser}\right) \quad , \qquad (3)$$

where ρ is the network utilization, W is the number of users, and $T_{tx}$ is the average time slot per user. This parameter is also known as network delay, and $T_{Laser}$ is the time duration served to each subscriber or user. By combining a strong CW pump signal at angular frequency ($\omega_p$) with a signal at another frequency ($\omega_s$) into a SMF or a HNLF cable, parametric gain can be achieved. At the same time, a converted signal, called idler ($\omega_i$), will be generated at the frequency $\omega_i = 2\omega_p - \omega_s$. The process is described using the following three coupled equations that describes the amplitudes $A_{p,s,i}$ of the pump, signal, and idler [13]:

$$\frac{dA_p}{dz} = i\gamma\left[\left(\left|A_p\right|^2 + 2\left(\left|A_s\right|^2 + \left|A_i\right|^2\right)\right)A_p + 2A_s\,A_i\,A_p^*\exp\left(i\,\Delta\beta\,z\right)\right], \qquad (4)$$

$$\frac{dA_s}{dz} = i\gamma\left[\left(\left|A_s\right|^2 + 2\left(\left|A_s\right|^2 + \left|A_i\right|^2\right)\right)A_s + A_i^*A_p^2\exp\left(-i\,\Delta\beta\,z\right)\right], \qquad (5)$$

$$\frac{dA_i}{dz} = i\gamma\left[\left(\left|A_i\right|^2 + 2\left(\left|A_s\right|^2 + \left|A_s\right|^2\right)\right)A_i + A_s^*A_p^2\exp\left(-i\Delta\beta\,z\right)\right], \qquad (6)$$

The equation for the pump amplitude can now be integrated to give:

$$A_p = A_{p_0}\ \exp\left(i\gamma\left|A_{p_0}\right|^2 z\right) = \sqrt{P_p}\,\exp\left(i\gamma P_p z\right) \quad , \qquad (7)$$

where $\left|A_{p_0}\right|^2 = P_p$ is the pump power at z = 0, which implies that the pump signal does not lose any power. In the no depletion approximation, the parametric amplification is described by the signal power gain [13] as:

$$G_s(L) = \frac{P_s(L)}{P_s(0)} = 1 + \left[\frac{\gamma P_p}{g}\sinh\left(gL\right)\right]^2 \quad , \qquad (8)$$

$$\gamma = \frac{2\,\pi\,n_2}{\lambda\,A_{eff}} \quad , \qquad (9)$$

where $P_p$, $P_s$ are the pump and signal powers in the fiber cable, γ is the fiber nonlinear coefficient, L is the fiber cable length, $A_{eff}$ is the effective cross-section area of the fiber, $n_2$ is the nonlinear refractive-index coefficient ≈ 2.6 $x10^{-20}$ $m^2/V^2$, λ is the optical signal wavelength, and g is the parametric gain parameter give by [13]:

$$g^2 = \left(\gamma P_p\right)^2 - k^2/4 = -\Delta\beta\left(\Delta\beta/4 + \gamma P_p\right), \qquad (10)$$

where k=Δ β+ γ $P_p$, and the propagation mismatch Δβ is given as follows [13]:

$$\Delta\beta = -\frac{2\pi c}{\lambda_0^2} S_p \left(\lambda_p - \lambda_0\right)\left(\lambda_p - \lambda_s\right)^2 , \qquad (11)$$

$$L_{eff} = \frac{1-\exp\left(-\alpha L\right)}{L} , \qquad (12)$$

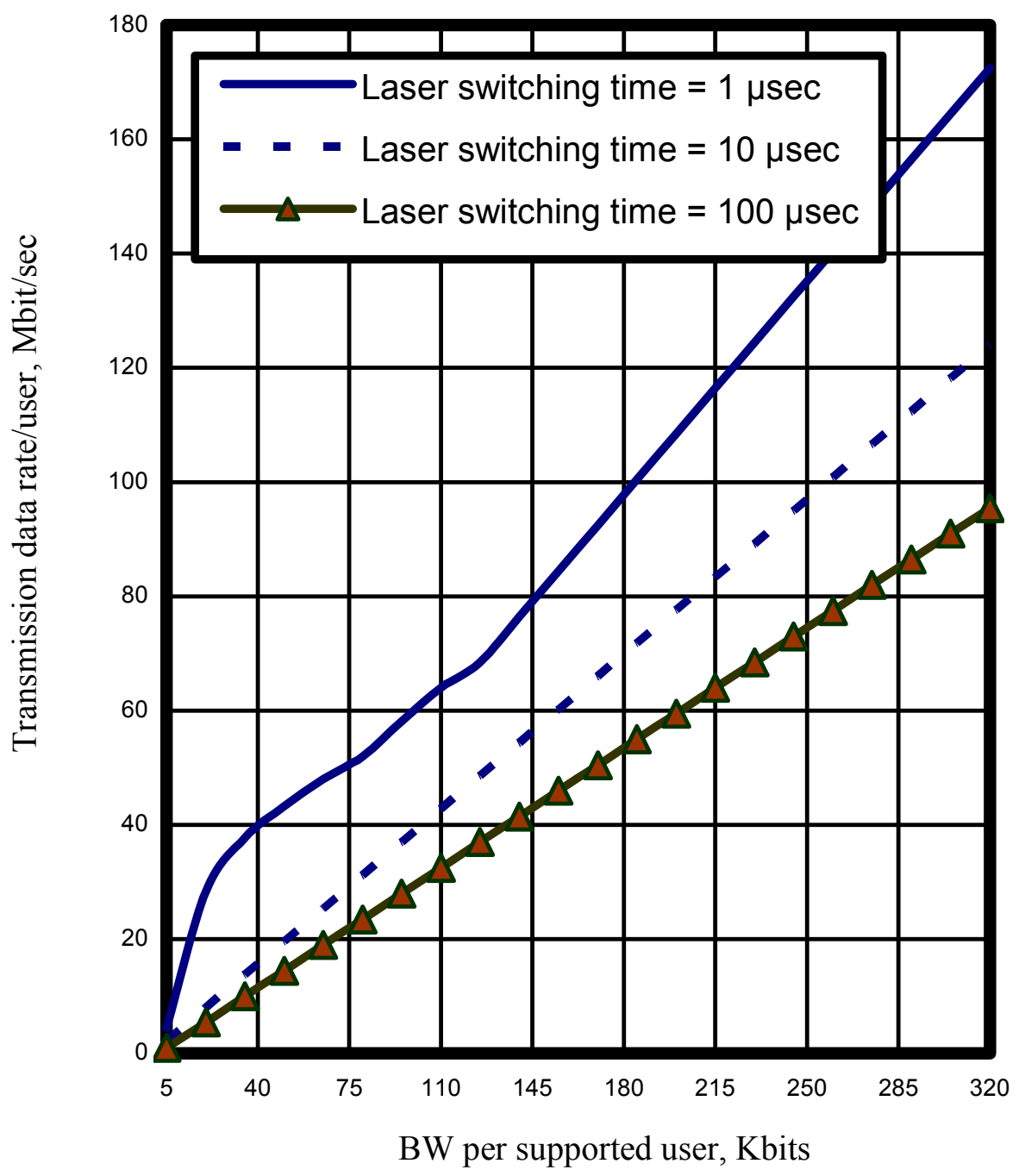

Figure 2. Variations of transmission data rate with BW per supported user at the assumed set of parameters.

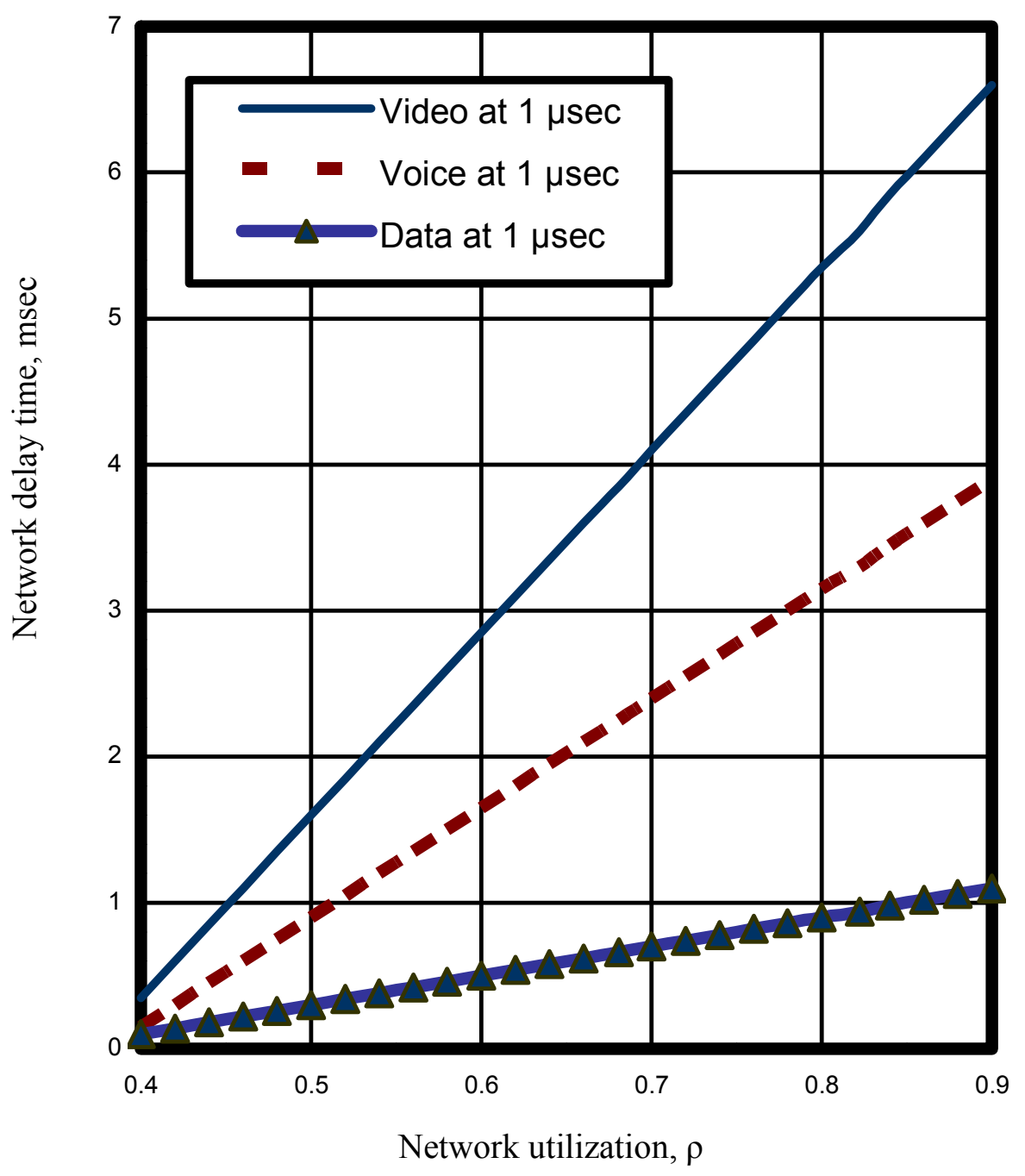

Figure 3. Variations of network delay time with the network utilization at the assumed set of parameters for three offered services.

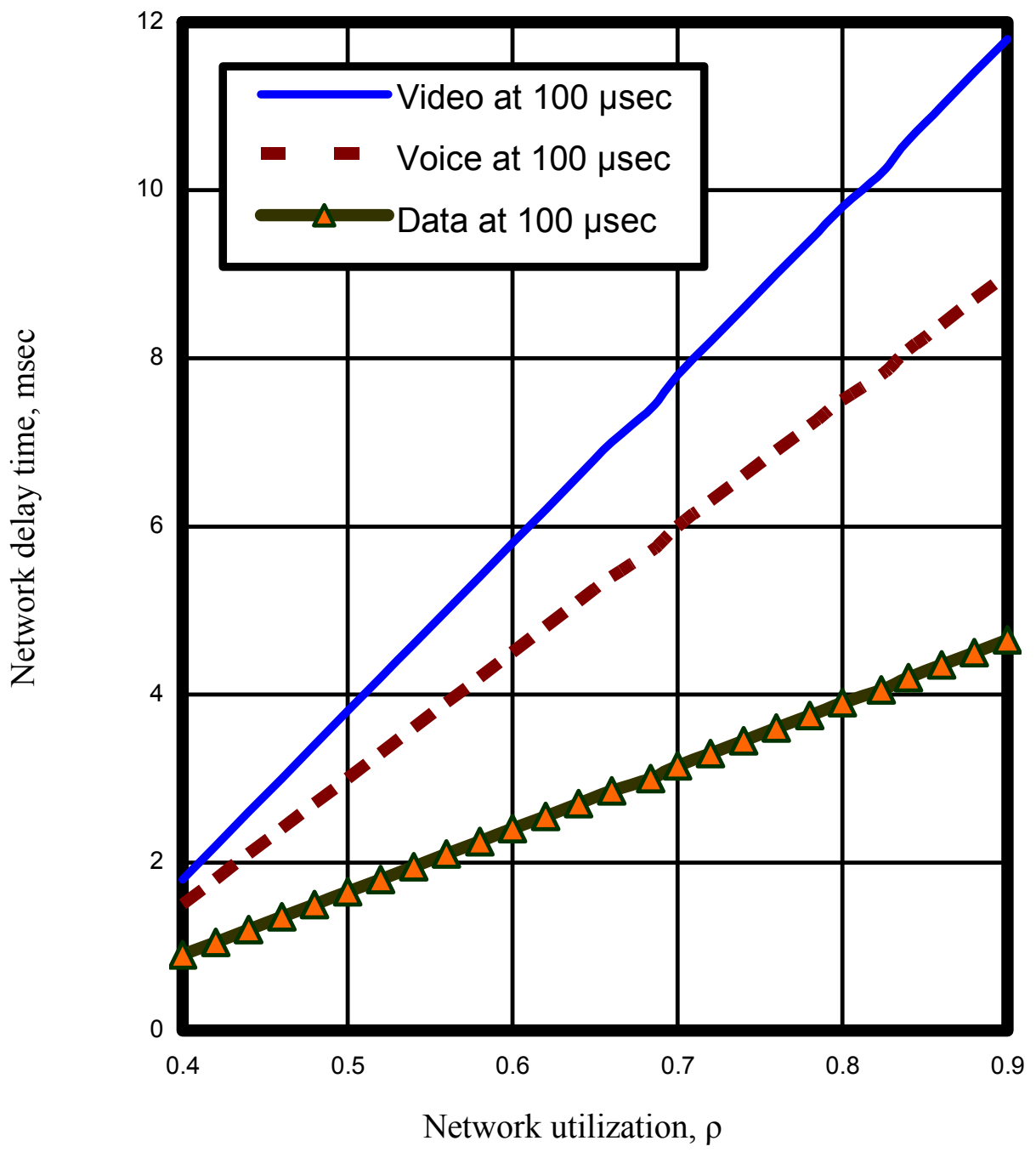

Figure 4. Variations of network delay time with the network utilization at the assumed set of parameters for three offered services.

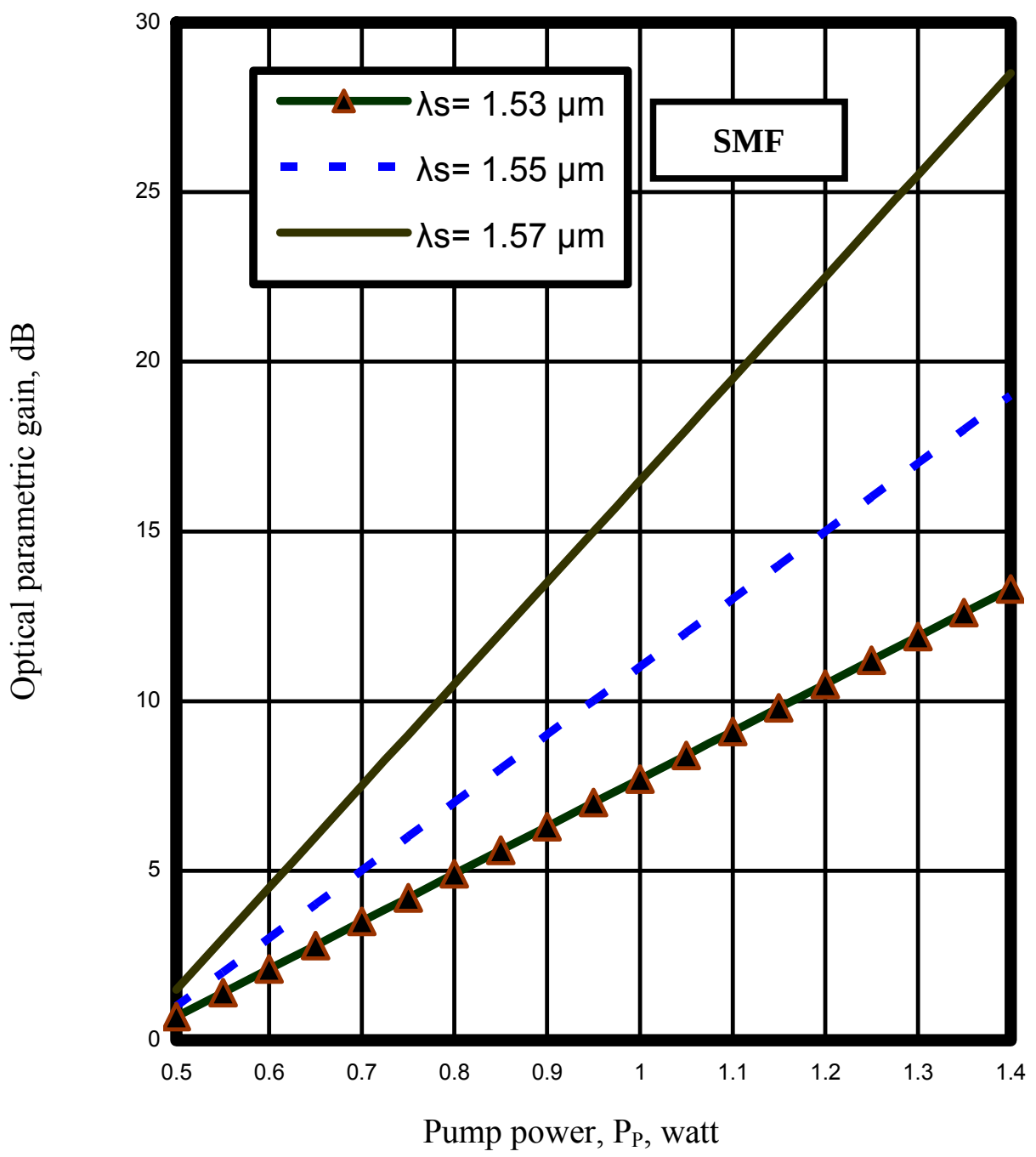

Figure 5. Variations of optical parametric gain with pump power at the assumed set of parameters.

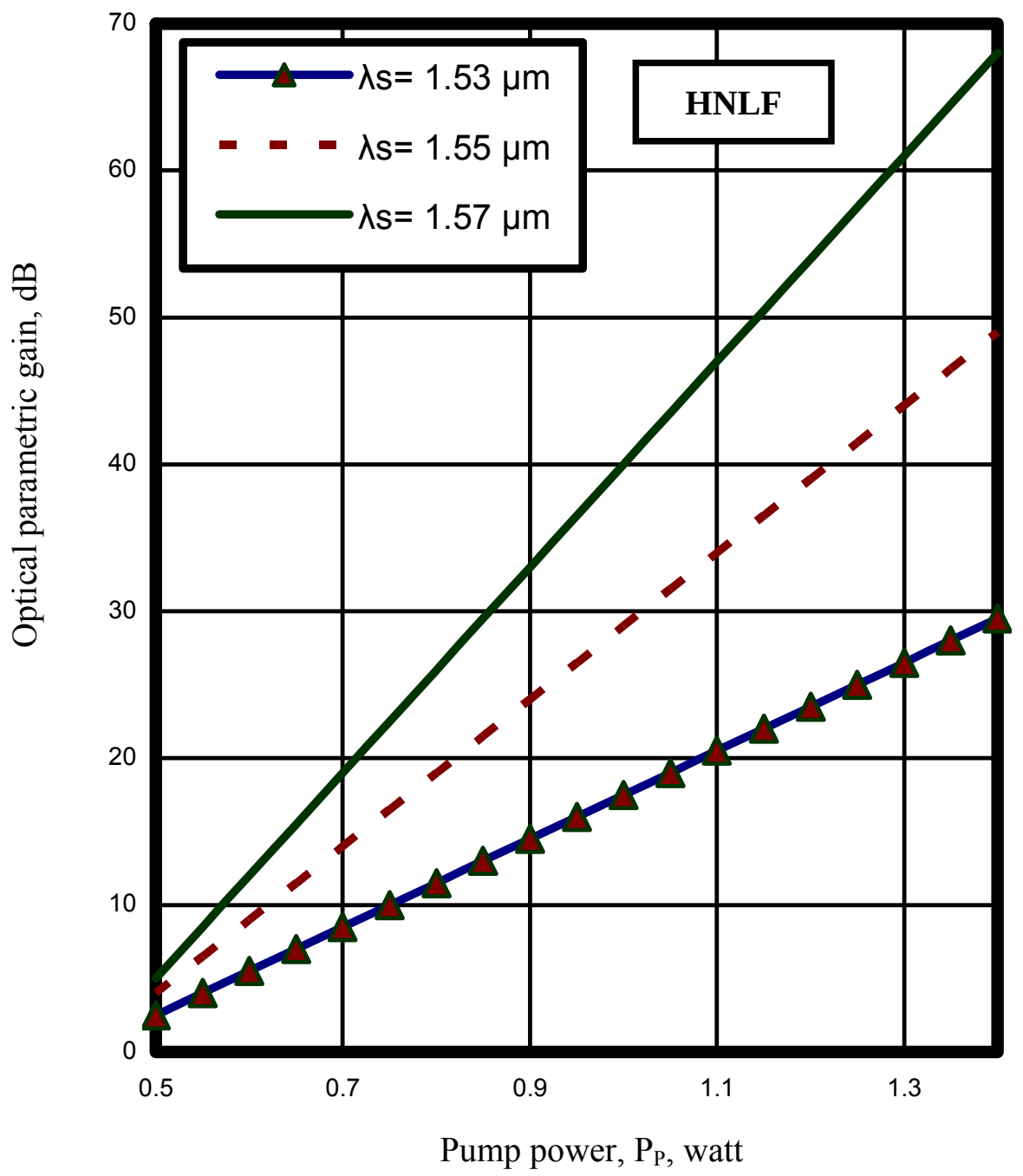

Figure 6. Variations of optical parametric gain with pump power at the assumed set of parameters.

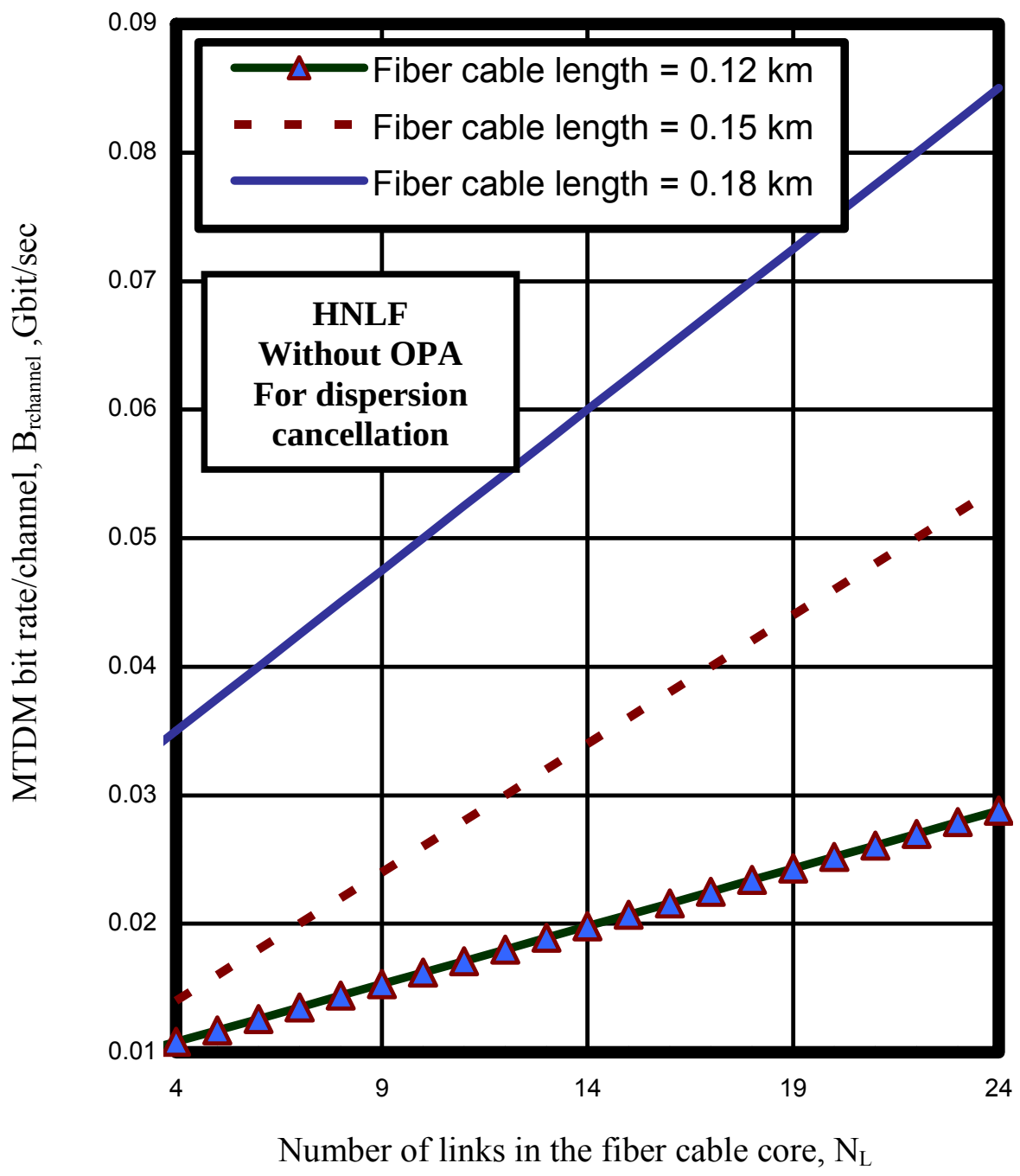

Figure 7. Variations of MTDM bit rate/channel with the number of links in the fiber cable core at the assumed set of parameters.

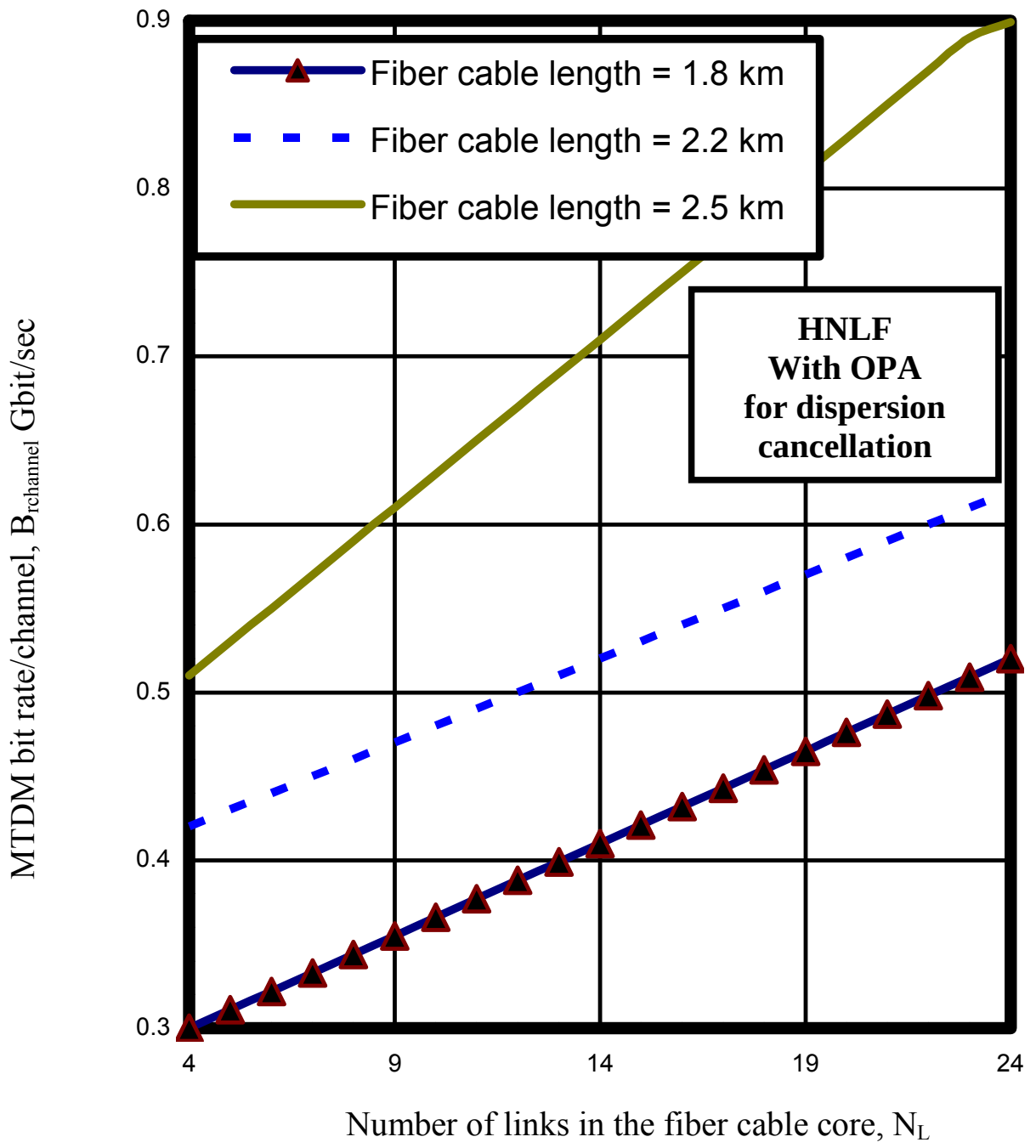

Figure 8. Variations of MTDM bit rate/channel with the number of links in the fiber cable core at the assumed set of parameters.

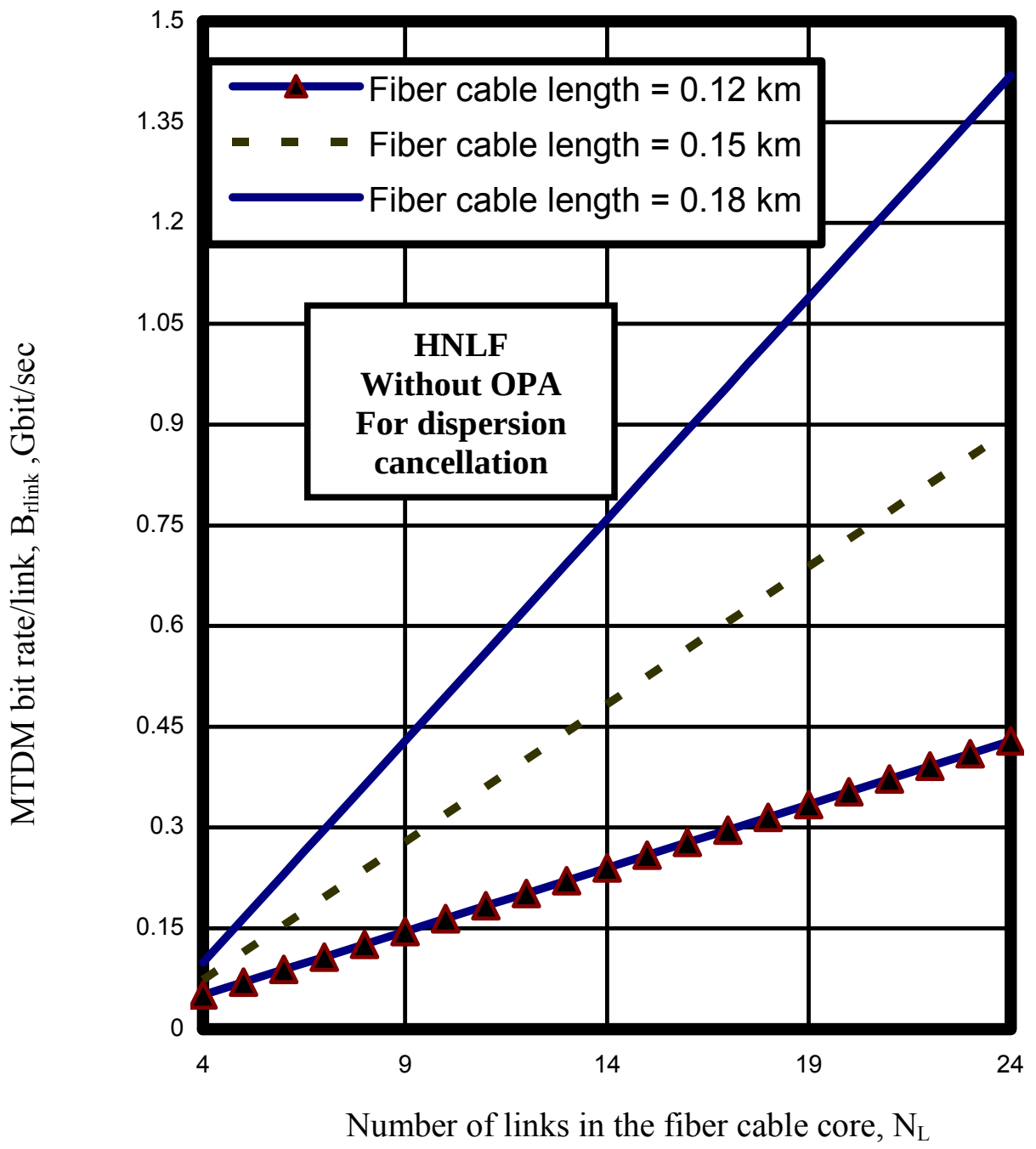

Figure 9. Variations of MTDM bit rate/link with the number of links in the fiber cable core at the assumed set of parameters.

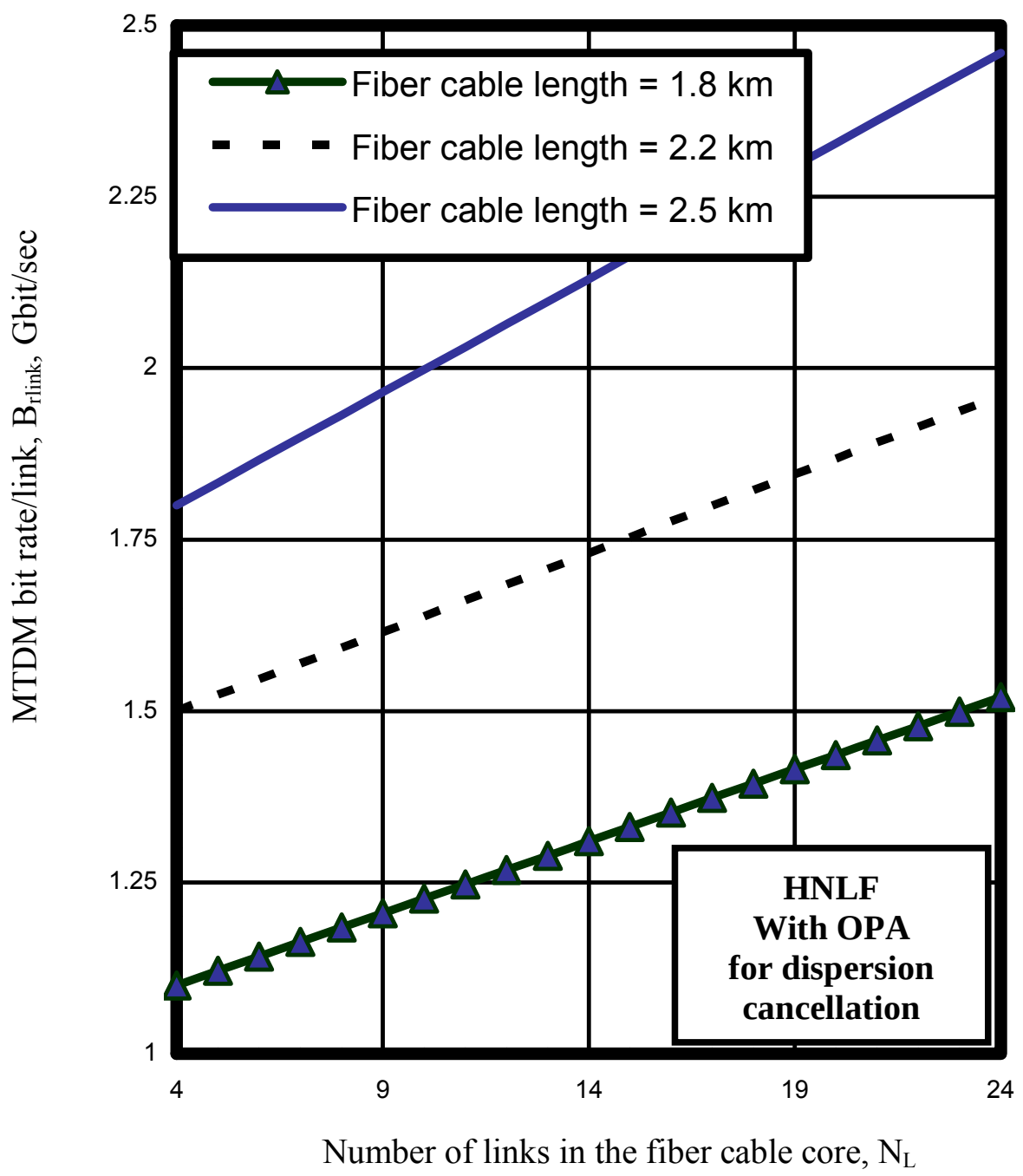

Figure 10. Variations of MTDM bit rate/link with the number of links in the fiber cable core at the assumed set of parameters.

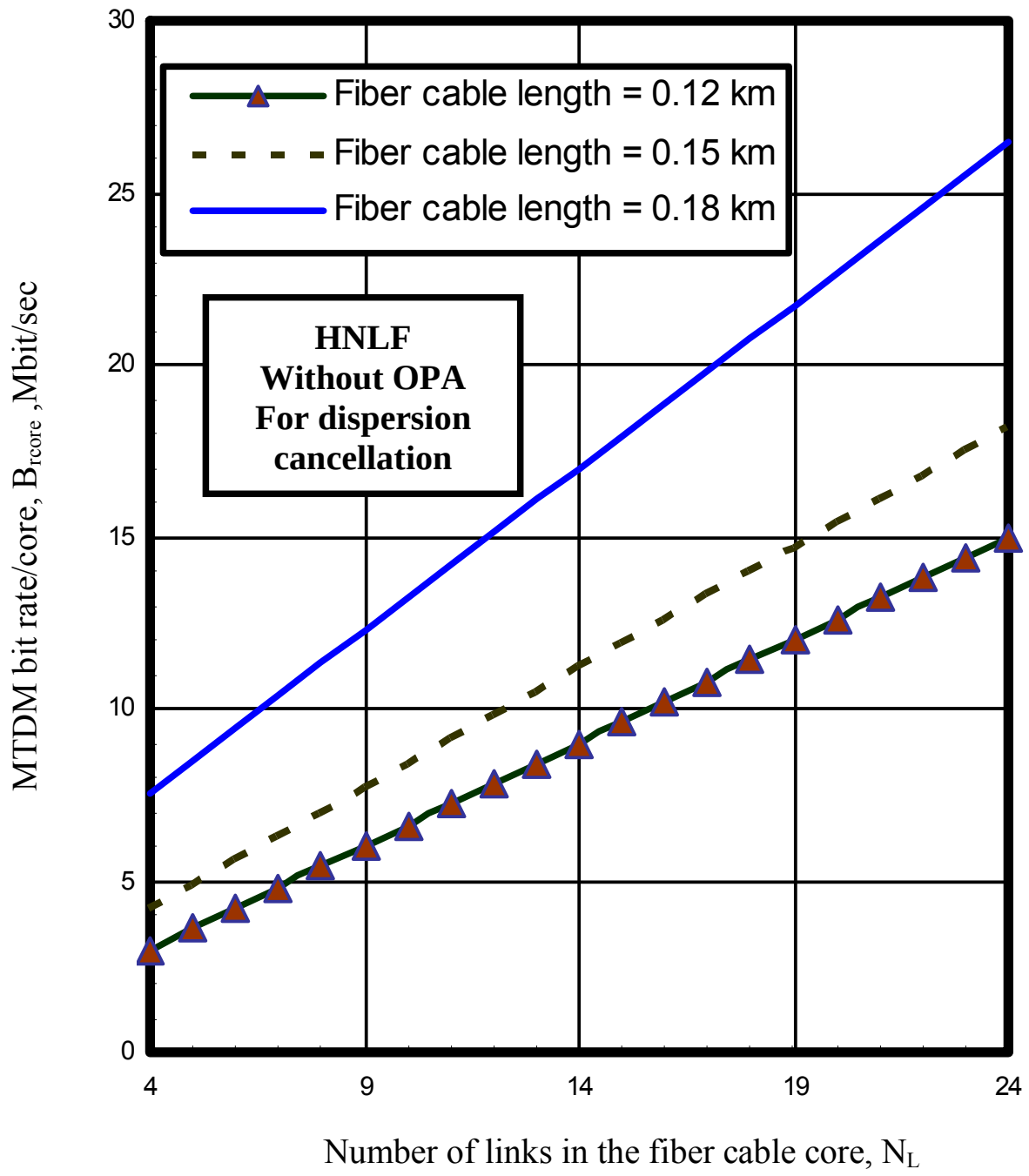

Figure 11. Variations of MTDM bit rate/fiber cable core with the number of links in the fiber cable core at the assumed set of parameters.

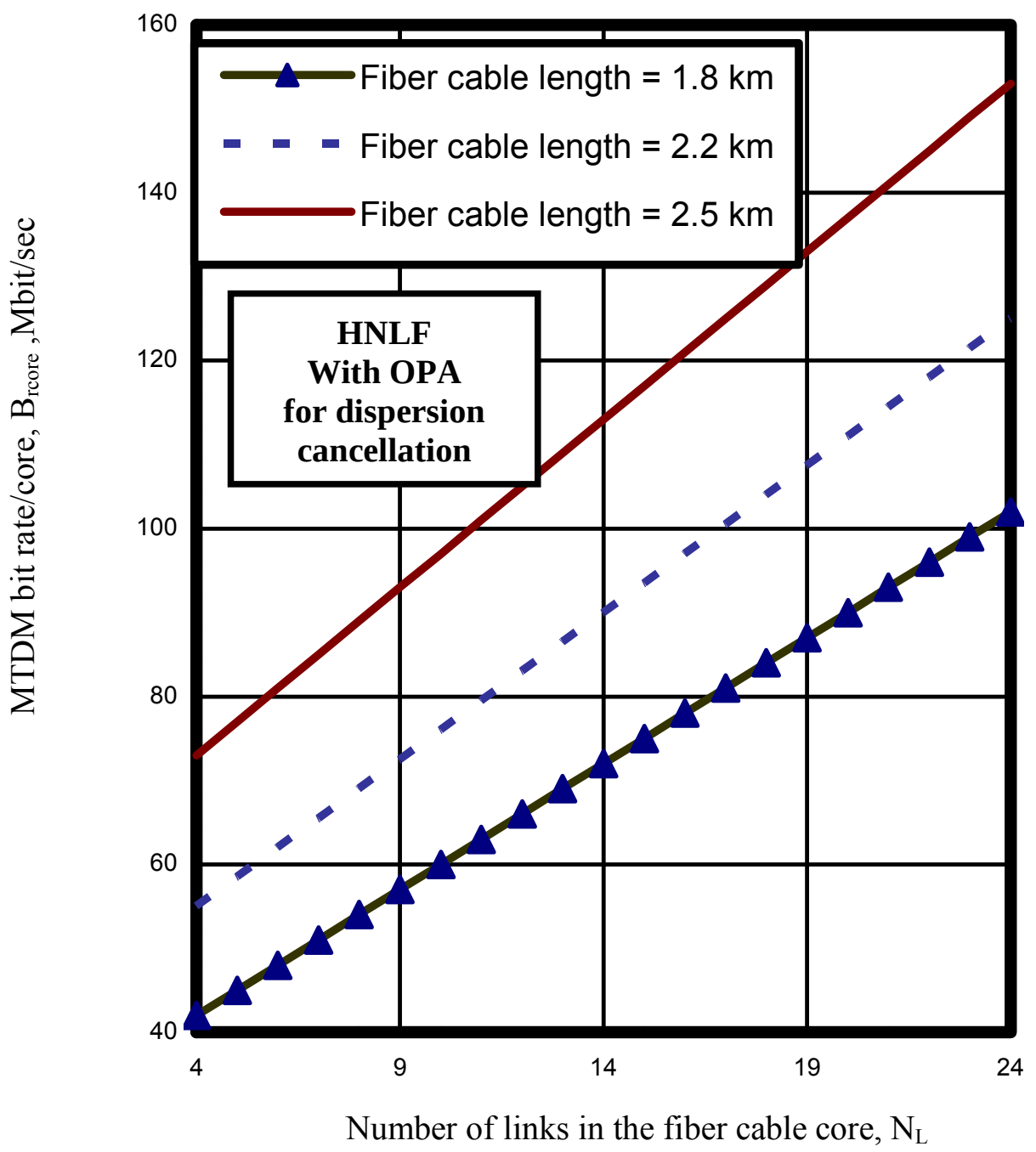

Figure 12. Variations of MTDM bit rate/fiber cable core with the number of links in the fiber cable core at the assumed set of parameters.

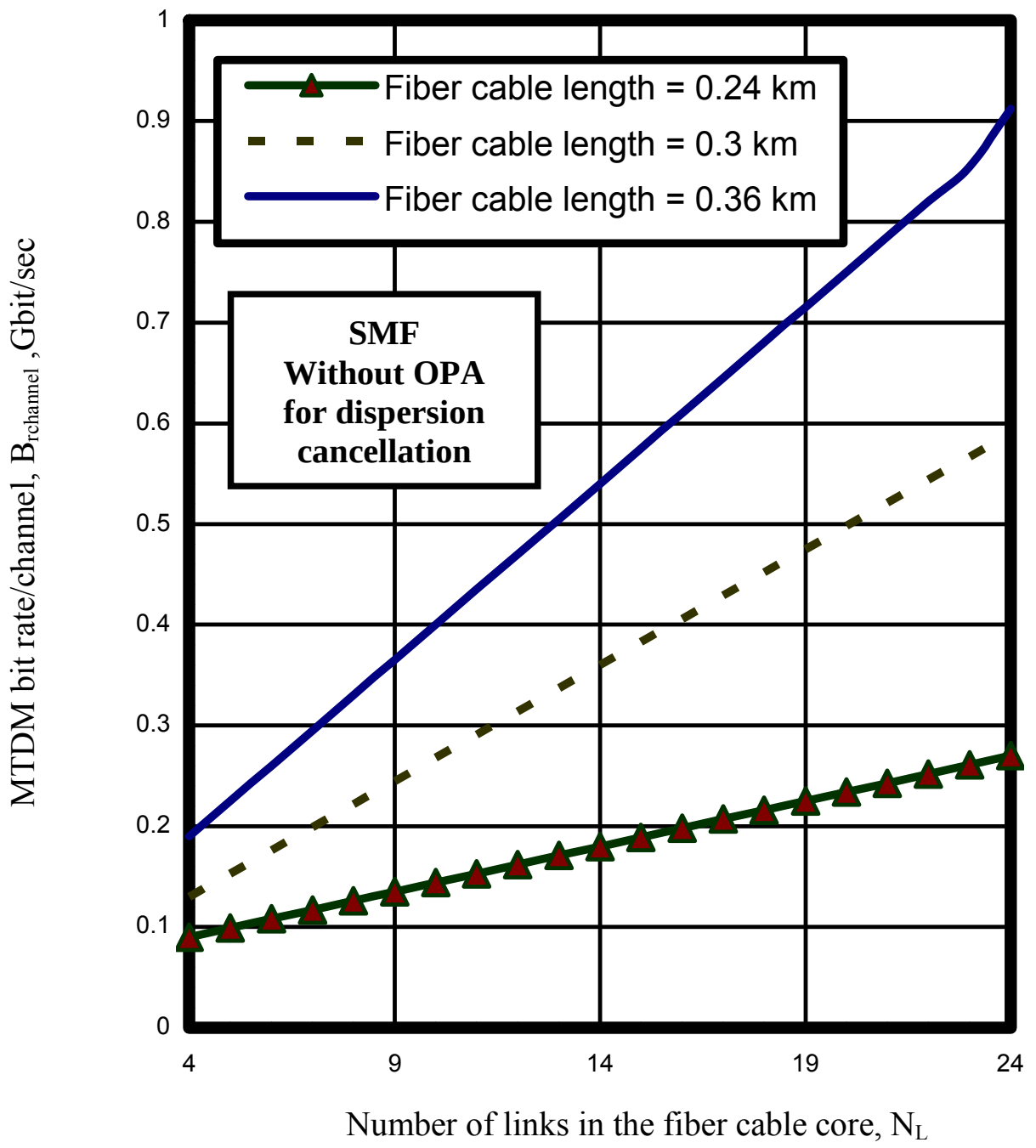

Figure 13. Variations of MTDM bit rate/channel with the number of links in the fiber cable core at the assumed set of parameters.

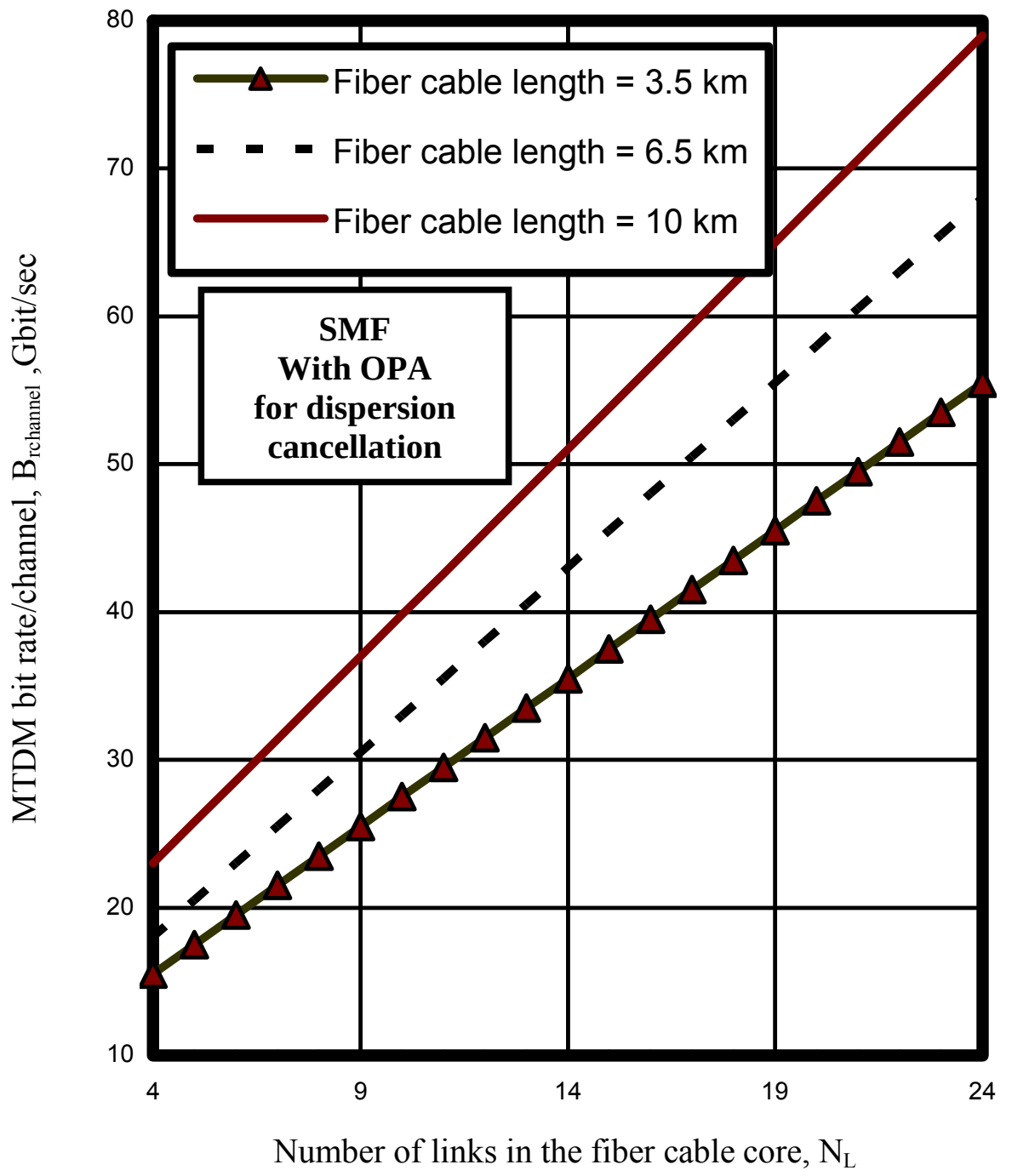

Figure 14. Variations of MTDM bit rate/channel with the number of links in the fiber cable core at the assumed set of parameters.

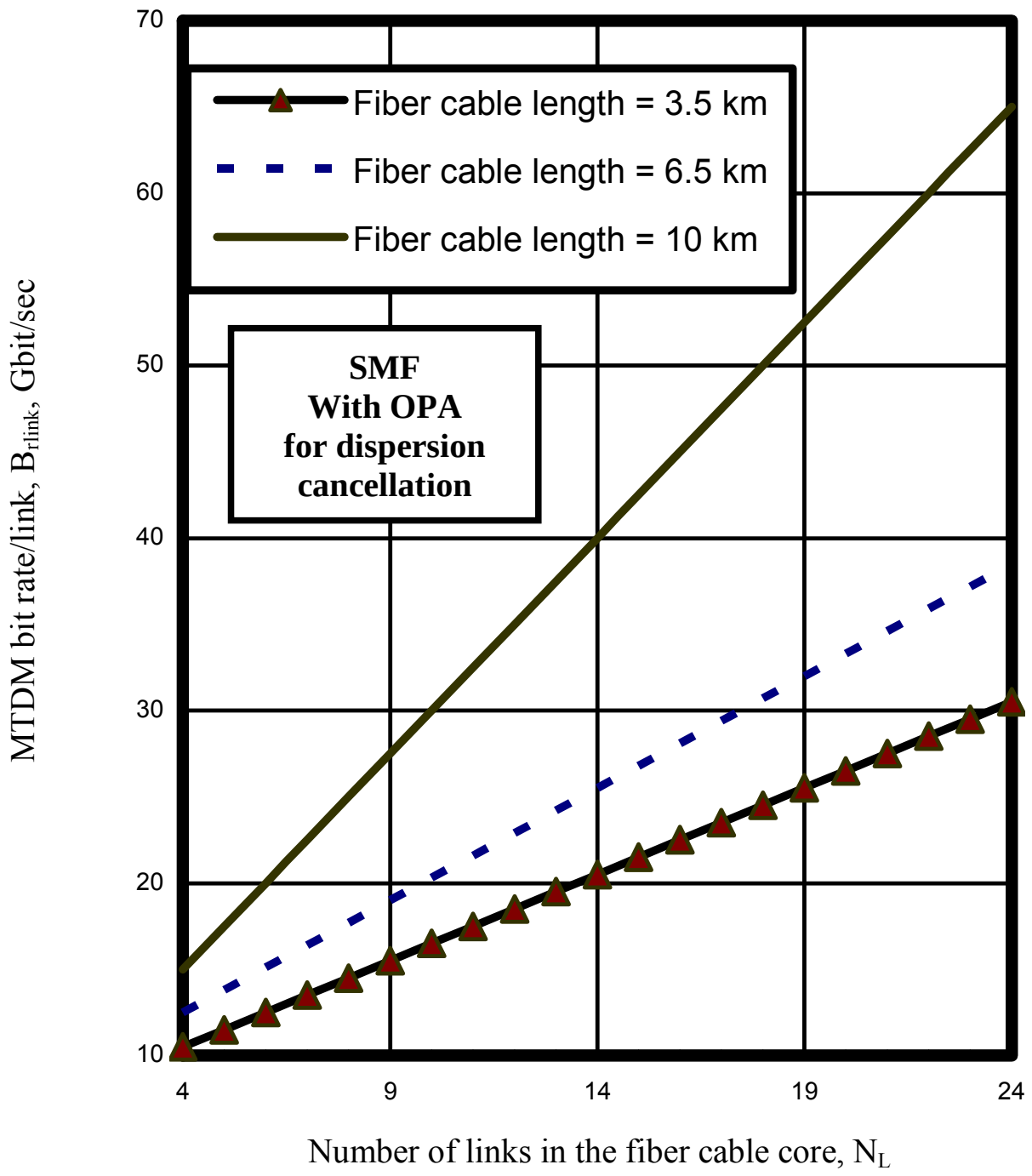

Figure 16. Variations of MTDM bit rate/link with the number of links in the fiber cable core at the assumed set of parameters.

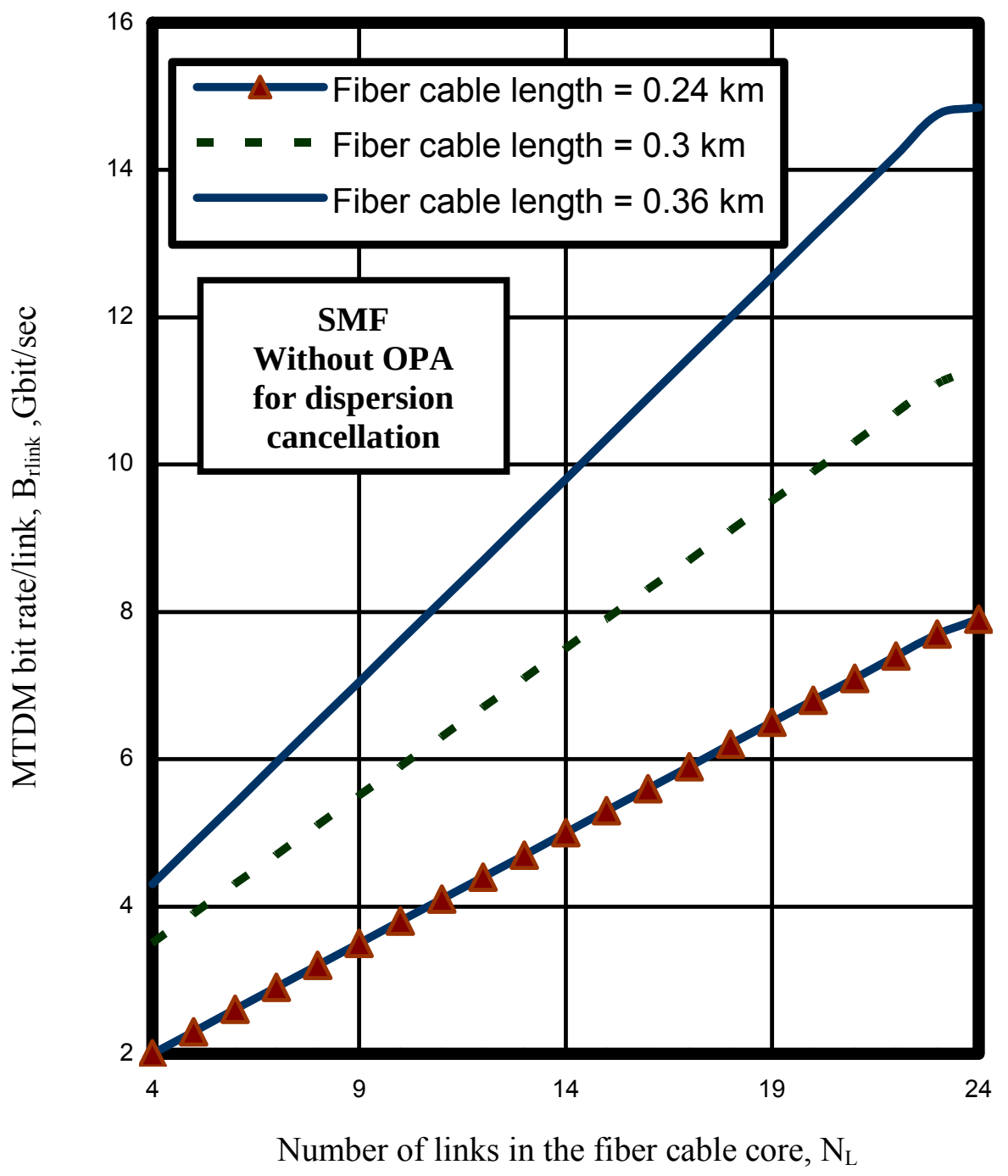

Figure 15. Variations of MTDM bit rate/link with the number of links in the fiber cable core at the assumed set of parameters.

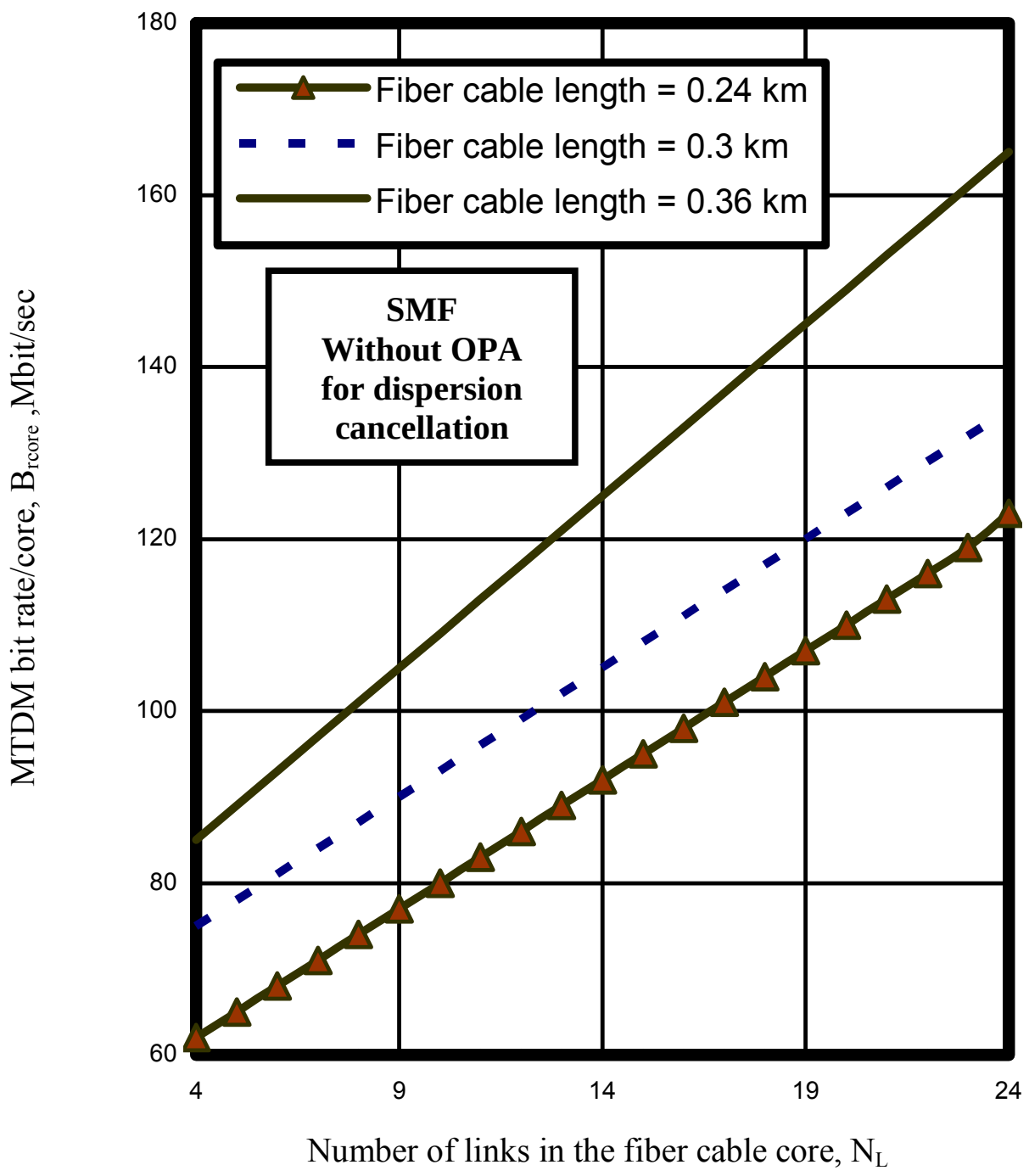

Figure17. Variations of MTDM bit rate/fiber cable core with the number of links in the fiber cable core at the assumed set of parameters.

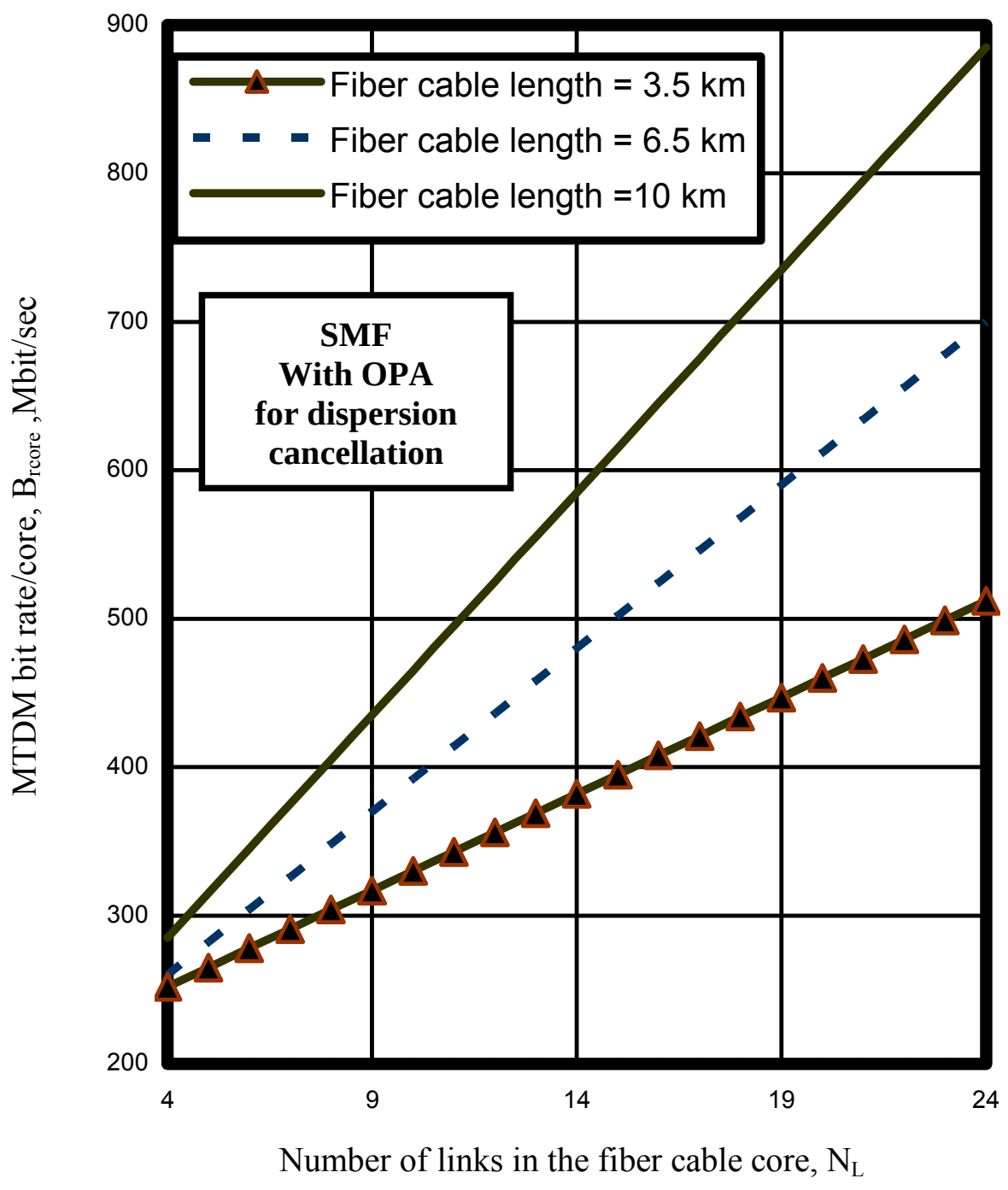

Figure 18. Variations of MTDM bit rate/fiber cable core with the number of links in the fiber cable core at the assumed set of parameters.

where $\lambda_0$ is the zero dispersion wavelength, $\lambda_p$ is the pump wavelength, $\lambda_s$ is the optical signal wavelength, c is the velocity of light, and $S_p$ is the parametric gain slope. If the fiber is long or the attenuation is high, the interaction length will be limited by the effective length $L_{eff}$ can be expressed as $L_{eff} \approx L$. The signal gain of the optical fiber parametric amplifier can be expressed as follows [14]:

$$G_s = 1 + (\gamma P_p)^2 \left[ 1 + \frac{gL^2}{6} + \frac{gL^4}{120} + ............ \right]^2 , \qquad (13)$$

From Eq. (13), it may be noted that for signal wavelengths close to $\lambda_p$, $\Delta\beta \approx 0$, and $G_s = (\gamma P_p L)^2$. In the special case of perfect phase matching ($k \approx 0$) and $\gamma P_p L$ »1, Eq. (13) can be rewritten as follows:

$$G_s = \sinh^2(gL) \approx \sinh^2(\gamma P_p L) \approx \frac{1}{4}\exp(2\gamma P_p L) \qquad (14)$$

A very simple expression for the OPA gain can be obtained if Eq. (14) is rewritten in decibel units as:

$$G_{dB} = 10\log_{10}\left[\frac{1}{4}\exp(2\gamma P_p L)\right] = P_p L S_p - 6 , \qquad (15)$$

$$S_p = 10\log_{10}[\exp(2)]\gamma \approx 8.7\gamma , \qquad (16)$$

where $S_p$ is introduced as the parametric gain slope in [dB/Watt. km]. Based on Eqs. (4-6), the derived expression for the generated pulse is obtained [13]. A sinusoidally modulated pump $P_p(t)$ is assumed together with a CW signal. The pump is also considered to be undeleted. They have found that the output pulses are approximately chirped Gaussian pulses in high gain regime defined as follows [13]:

$$A(0,t) = A_0 \exp\left[ -\frac{1+iC}{2}\left(\frac{t}{T_0}\right)^2 \right] , \qquad (17)$$

where the amplitude $A_0$ can be identified as follows [14]:

$$A_0 = \frac{2\exp(g_0 L)}{2 g_0} , \qquad (18)$$

where $g_0 = g$ (t=0) is given by the following expression:

$$g_0 = \frac{\sqrt{-\Delta\beta^2 + 4\Delta\beta\gamma P_0}}{2} , \qquad (19)$$

and the pulse width duration $T_0$ is given by the following expression [14]:

$$T_0 = \sqrt{\frac{2 g_0}{\Delta\beta\gamma P_0'' L}} , \text{ nsec} \qquad (20)$$

In these expression $P_0 = P_P(0)$ is the peak pump power and $P_0''$ is the second derivative with respect to time of $P_P$ at the peak value. The pulse width $T_0^2$ is proportional to 1/L. The optical signal wavelength span 1.5 μm ≤ $\lambda_{si}$, optical signal wavelength≤ 1.65 μm is divided into intervals per link as follows:

$$\Delta\lambda_0 = \frac{\lambda_f - \lambda_i}{N_L} = \frac{0.15}{N_L} , \ \mu m/link \qquad (21)$$

The transmitted MTDM bit rates per optical network channel is computed as follows [15]:

$$B_{rchannel} = \frac{1}{4T_0} = \frac{0.25}{T_0} , Gbit/\sec/channel \qquad (22)$$

Then the MTDM bit rates per fiber cable link is given by the following expression:

$$B_{rLink} = \frac{0.25 \ x N_{ch}}{T_0} , Gbit/\sec/link \qquad (23)$$

Therefore, the total MTDM bit rates per fiber cable core is given by the following expression:

$$B_{rCore} = \frac{0.25 x 1000 \ x N_L \ x N_{ch}}{T_0} , Mbit/\sec/core \qquad (24)$$

where $N_{ch}$ is the number of optical network channels in the fiber cable link, and $N_L$ is the number of links in the fiber cable core.

## IV. RESULTS AND DISCUSSIONS

We have investigated the basic MTDM transmission technique to transmit multi-optical network channels with higher bit rates based on both WDM, and TDM With the assistant of OPAs in the interval of 1.5 μm to 1.65 μm. The following numerical data (set of the controlling parameters) of our system model are employed to obtain the performance of hybrid WDM/TDM local area PON with the assistant of OPAs: 1.5 ≤ $\lambda_{si}$, optical signal wavelength, μm ≤ 1.65, 1.4 ≤ $\lambda_p$, pumping wavelength, μm ≤ 1.55, 0.5 ≤ $P_P$, pump power, Watt/pump ≥ 1.4, $N_L$: total number of links up to 24 links, number of laser diodes: K= 16 lasers, number of input ports of AWG Mux: M=16 channels, number of output ports: N= 16 channels, number of users: W=256 user, and the fiber cable parameters for different fiber cable types as shown in Table 1.

TABLE 1. PHYSICAL PARAMETERS ARE USED IN PROPOSED HYBRID NETWORK MODEL FOR DIFFERENT FIBER CABLE TYPES [13].

| Fiber Parameters | SMF | HNLF |
|---|---|---|
| Attenuation, $\alpha$ [dB/km] | 0.2 dB/km | 0.7 dB/km |
| Effective area, $A_{eff}$ [$\mu$m] | 85 $\mu$m | 12 $\mu$m |
| Nonlinear coefficient at 1.55 $\mu$m, $\gamma$ [$Watt^{-1}$. $km^{-1}$] | 1.8 $Watt^{-1}$. $km^{-1}$ | 15 $Watt^{-1}$. $km^{-1}$ |
| Parametric gain slope, $S_p$ [dB/Watt.km] | 16 | 131 |

Based on the above governing equations analysis of the proposed hybrid network model, the set assumed of controlling data parameters, and the series of the Figs. (2-18), the following features are assured:

1) Figure 2 has indicated that the bandwidth per supported user increases, the transmission data rate per user also increases at the same laser switching time. While, as the laser switching time increases, the transmission data rate per user decreases at the same bandwidth per supported user.
2) Figs. (3, 4) have demonstrated that as the network utilization increases, the network delay time also increases at the same laser switching time. Moreover, as the laser switching time increases for three offered services, the network delay time also increases at the same network utilization.
3) As shown in Figs. (5, 6), as the pump power increases, the optical parametric gain also increases across both HNLF and SMF cables at the same optical signal wavelength. Moreover, as the optical signal wavelength increases, the optical parametric gain also increases across both HNLF and SMF cables at the same pump power.
4) Figs. (7, 8) have explained that as the number of links in the fiber cable core increases, the MTDM bit rate per channel also increases at the same fiber cable length. Moreover, as the fiber cable length increases. The MTDM bit rate per channel also increases at the same number of links. But we observed that with OPA across HNLF cable gives long distance transmission for network reach, and higher bit rates per channels.
5) Figs. (9, 10) have demonstrated that as the number of links in the fiber cable core increases, the MTDM bit rate per link also increases at the same fiber cable length. Moreover, as the fiber cable length increases. The MTDM bit rate per link also increases at the same number of links. But we observed that with OPA across HNLF cable gives long distance transmission for network reach, and higher bit rates per links.
6) Figs. (11, 12) have indicated that as the number of links in the fiber cable core increases, the MTDM bit rate per fiber core also increases at the same fiber cable length. Moreover, as the fiber cable length increases. The MTDM bit rate per fiber core also increases at the same number of links. But we observed that with OPA across HNLF cable gives long distance transmission for network reach, and higher bit rates per fiber core.
7) Figs. (13, 14) have given that as the number of links in the fiber cable core increases, the MTDM bit rate per channel also increases at the same fiber cable length. Moreover, as the fiber cable length increases. The MTDM bit rate per channel also increases at the same number of links. But we observed that with OPA across SMF cable gives long distance transmission for network reach, and higher bit rates per channels.
8) Figs. (15, 16) have shown that as the number of links in the fiber cable core increases, the MTDM bit rate per link also increases at the same fiber cable length. Moreover, as the fiber cable length increases. The MTDM bit rate per link also increases at the same number of links. But we observed that with OPA across SMF cable gives long distance transmission for network reach, and higher bit rates per links.
9) As shown in Figs. (17, 18), as the number of links in the fiber cable core increases, the MTDM bit rate per fiber core also increases at the same fiber cable length. Moreover, as the fiber cable length increases. The MTDM bit rate per fiber core also increases at the same number of links. But we observed that with OPA across SMF cable gives long distance transmission for network reach, and higher bit rates per fiber core.

## V. CONCLUSIONS

In a summary, the OPAs are employed over wide range of the affecting parameters in hybrid WDM/TDM local area Passive optical networks across SMF, or HNLF cables to achieve the best QOS to handle a triple play solution to the supported users. We have demonstrated that the fast of the laser switching time, the higher of transmission data rate per supported user, and the lower network delay time to handle the offered services as voice, video, and data for the supported users. Moreover, we have demonstrated that the higher of both pumping power, and optical signal wavelength, the higher optical parametric gain across HNLF, and SMF cables. It is evident that the OPAs play a vital role in extended network reach with higher bit rates either per links or per channels across both HNLF and SMF cables. Finally, we have demonstrated that OPAs with SMF cables have offered higher bit rates either per link or per channels than HNLF cables for the same extended network reach.

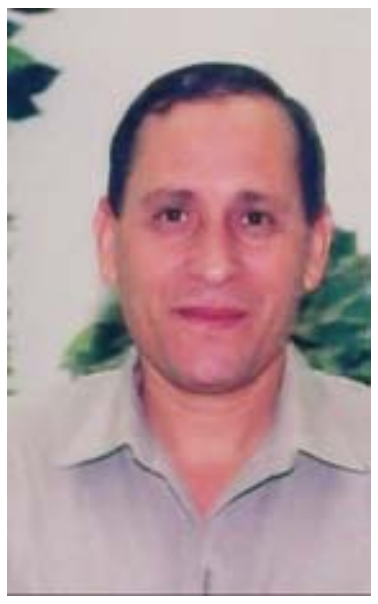

**Abd-Elnaser A. Mohammed**

Received Ph.D scientific degree from the faculty of Electronic Engineering, Menoufia University in 1994. Now, his job career is Assoc. Prof. Dr. in Electronics and Electrical Communication Engineering department. Currently, his field and research interest in the all passive optical and communication Networks, analog-digital communication systems, optical systems, and advanced optical communication networks.

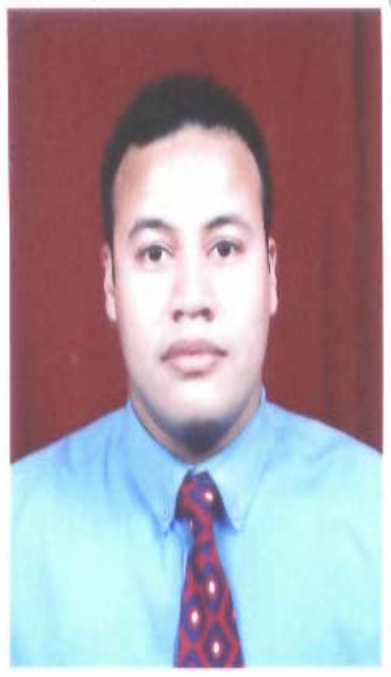

**Ahmed Nabih Zaki Rashed**

was born in Menouf, Menoufia State, Egypt, in 1976. Received the B.Sc. and M.Sc. practical scientific degrees in the Electronics and Electrical Communication Engineering Department from Faculty of Electronic Engineering, Menoufia University in 1999 and 2005, respectively. Currently, his field interest and working toward the Ph.D degree in Active and Passive Optical Networks (PONs). His theoretical and practical scientific research mainly focuses on the transmission data rates and distance of optical access networks.

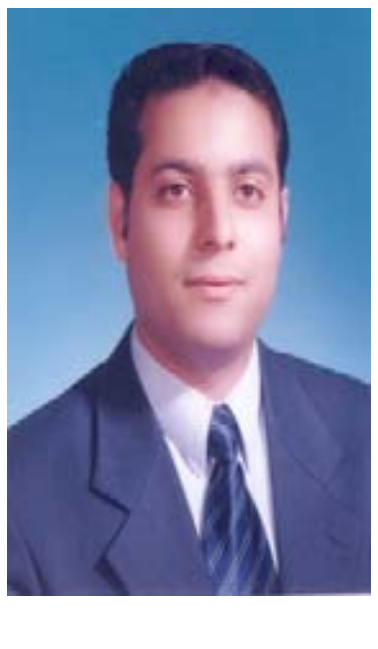

**Mohamoud M. A. Eid**

was born in gharbiya State, Egypt, in 1977. Received the B.Sc. and M.Sc. degrees in the Electronics Communication Engineering Department from Faculty of Electronic

Engineering, Menoufia University in 2002 and 2007. Currently, his working toward the Ph.D degree in Ultra wide wavelength division multiplexing.